# Software Defect Prediction Using Support Vector Machine

Haneen Abu Alhija, Mohammad Azzeh, and Fadi Almasalha

Faculty of Information Technology, Applied Science Private University, Amman, Jordan, 11391

h_Abualhija@asu.edu.jo , m.y.azzeh@asu.edu.jo ,

**Abstract** Software defect prediction is an essential task during the software development Lifecycle as it can help managers to identify the most defect-proneness modules. Thus, it can reduce the test cost and assign testing resources efficiently. Many classification methods can be used to determine if the software is defective or not. Support Vector Machine (SVM) has not been used extensively for such problems because of its instability when applied on different datasets and parameter settings. The main parameter that influences the accuracy is the choice of the kernel function. The use of kernel functions has not been studied thoroughly in previous papers. Therefore, this research examines the performance and accuracy of SVM with six different kernel functions. Various public datasets from the PROMISE project empirically validate our hypothesis. The results demonstrate that no kernel function can give stable performance across different experimental settings. In addition, the use of PCA as a feature reduction algorithm shows slight accuracy improvement over some datasets.

## 1. Introduction

Predicting defect-prone modules during the software development process is crucial because it helps the quality assurance team put more effort into modules with a high probability of defect-proneness. It also helps the management team assign and distribute resources efficiently during testing, thus reducing development costs [1] [2]. The process of manually reviewing the code usually leads to a detection rate between 35% - 60% in most cases, but this rate is increased when defect prediction tools are used. Furthermore, the time needed to detect defect-proneness modules is reduced [3].

Software Defect Prediction (SDP) is performed by extracting static code metrics from bug log files of previous versions of the program, then using these static metrics for building models to predict the possible defects in future releases of the program [1] [4]. This process helps detect the location of the parts of the program that are likely to induce defects. It is used in a software system with a limited project budget or too large to be tested exhaustively. SDP can be primarily used in two ways: within a project or cross-project. The first approach implies using the same data as training and testing during the empirical validation process. In the second approach, one release of the project data is used as training, and the subsequent release is used as testing. Both approaches are acceptable and depend on data availability [5].

Any SDP model comprises four main elements:1) independent features representing static code metrics, 2) output features representing the presence of a defect or its absence. 3) The learning approach, and finally 4) the performance measures that are used to judge the accuracy of the built learning model [6]. The current studies on SDP models focus on four research aspects. The first aspect examines the importance of static code metrics for defect prediction and which metrics are more predictive than others [5] [7] [8] [9]. The second aspect focus on building defect prediction models from within data or across data [10]. The third aspect studies the effect of imbalanced data on the accuracy of defect prediction models [1] [11] [12] [13]. Finally, the fourth aspect focuses on using ranking techniques to predict the correct rank of the defected modules based on their number of defects [14] [4]. This study focuses on the second research aspect and attempts to study the performance of support vector machine with different kernel techniques for software defect prediction problems. Support Vector Machine (SVM) is an efficient machine leaning method that is suited for classification problems, as in our case [15][16][17]. The SVM has not been studied thoroughly in previous papers because of the instability of its accuracy over multiple datasets and it is easily influence by the choice of kernel functions [18][19][20]. This study attempts to bridge that research gap. Different Kernel functions will be used to test the accuracy of SVM for defect prediction problems [21][22]. This paper aims to study the impact of different Kernel functions in support vector machine for the problem of software defect prediction. Six public datasets will be used to empirically validate and test our hypothesis and assumptions. These datasets were obtained from PROMISE data repository.

The rest of paper is structured as follows: Section 2 presents related work. Section 3 presents datasets. Section 4 shows methodology of our research. Section 5 presents results and discussion. Finally, section 6 presents conclusion.

## 2. Related Work

Sheppard et al. [12] examined the factors that affect the prediction of software defects. 42 studies out of 600 studies were used for the meta-analysis. The challenges were examined by the NOVA model so that the prediction process was divided into the groups: (1) Classifier family: in this group, the defect prediction techniques were divided to 7 main sections; Decision Tree, Recognition, SVM, Neural Network (ANN), Naive Bayes, CBR, Search and Benchmark. (2) Data set family: In this group, the Dataset had been divided into 24 this group, the Dataset had been divided into change or static metrics. (4) Researcher Group: There are two clusters of researchers; the most significant cluster is 8-10 researchers. The meta-analysis revealed strong evidence that current experiments in predicting defects are inadequate and ineffective. Okutan et al. [9] used Bayesian Network to find the relationships among metrics and defect proneness in different datasets. The PROMISE data repository used many public datasets for this experiment, such as Ant, Tomcat, Jedit, Velocity, Synapse, Poi, Lucene, Xalan, and Ivy. The static metrics used were LOC, CBO, LOCQ, WMC, RFC, LCOM, LCOM3, DIT, and NOC. Each of these datasets has a version number and instance number. The results show that the Lack of Coding Quality (LOCQ) has been evaluated as one of the best scores in the experiments.

Son et al. [10] studied the prediction of software defects through systematic mapping and establishing a protocol for the mapping. The processes of systematic mapping have been done in four stages :(1) Application of Inclusion-Exclusion Criteria: this stage is divided into two stages Inclusion Criteria and Exclusion Criteria. Inclusion Criteria: In this stage of the study, the defect predication used software metrics provided by the analysis of statistical, search-based and machine learning techniques. Exclusion Criteria: In this stage study, the defect prediction does not use dependent variable and non-empirical nature. (2) Quality Analysis: at this stage, choose the evaluation methodology used (3) Data Extraction: what kind of data is used (4) Data Synthesis: involves the accumulation of facts from the collected data during the data extraction process. This experiment used the techniques: Decision Tree, Support Vector Machine, Neural Network, Regression and Bayesian Learning. The Dataset was taken from different resources such as NASA, Eclipse, Mozilla, etc. The result indicated good accuracy when using a large Dataset with different metrics.

He et al. [8] used the Dataset from PROMISE, they selected 34 releases; each one has a number of instances files and number of defects. This study used 1) several independent variables which represent the inputs that will affect the dependent variable. The study used 20 static code metrics including CK suite, Martin's metrics, QMOOM suite, Extended CK suite, McCabe's CC, and LOC. 2) Dependent variables: which represents the outputs and effect, it was studied to see how much it varies as the independent variables change. It used different machine learning algorithm; in order to evaluate the result such as J48, Decision Tree, Support Vector Machine, Logistic Regression, and Naïve Bayes. The result showed that the simple metrics could be helpful to predict software defect.

Yang et al. [4] proposed a new approach of learning to rank using the rank task. The study used 11 different types of Dataset such as Eclipse, Lucene, Mylyn, PDA and other data. The study used different method (RF, RP, BART, NBR, ZINBR, ZIPR, HNBR, and HPR) to Compare the results for the 11-datasetsusing three different metrics. The study used 10 Cross-Validation. The result showed two benefits (1) learning to rank just do rank defects and does not need to predict defects for each module (2) these expected numbers were used to predict which modules are more flawed than others in project. Wang et al. [1] examined the problem of imbalance distribution, which may be a problem or can help to predict defect in software; through using 10 datasets from PROMISE; each one of these datasets has different number of features, different language and has a different percent of defect. This Dataset uses in different techniques in two top-ranked predictors machine learning. Naive Bayes and Random Forest and compare the result with other techniques PD, PF, balance, G-mean and AUC. The result showed that the balance and G-mean is the best result, which mean that it could use the imbalance distribution to help in predict defect.

Hassan [15] used predict the defect of program based on the change cod of complexity. There are many processes that can be associated with code change, including the pattern of source code modification, recorded by the source control systems, and a log that saves all dates that have been changed. Statistical Linear Regression (SLR Model) was built to predict faults in subsystem. Different models and different application were used. The result showed that complex code change process negatively affects the software system, and the more complex changes to a file, the higher the chance the file will contain fault.

## 3. Datasets

To evaluate the effectiveness of defect prediction, we are conducting experiments on a set of data available on the PROMISE website and which have been collecting data from NASA. The data from NASA come from different project. These public datasets include the information on space craft instrumentation, satellite flight control, and ground data for storage management. In this research we will use six public datasets that are most widely used in among researchers from this repository (CM1, JM1, KC1, PC1, Class-level data for KC1version 1 and Class-level data for KC1 Version 2). Each of these datasets possesses several software modules with input as the quality metrics. the outputs of each models are whether the program is defective or non- defective. The features are divided into two main parts: McCabe and Halstead measure. This measure defines "modules" as the smallest functional units. All these datasets were developed in either C or C++ language as shown in Table 1. From Table 2, can be noted that, for all the considered six datasets, JM1, CM1, KC1 and PC1 have 22 attributes. Each of this Dataset have been including one output attributes which represent the goal of filed (defect as 1, non-defect as 0) other attributes represent the quality metrics for the project acting as input attributes. These attributes can be classified in to McCabe metrics, 9 Halstead measures, and 8 are derived Halstead measures.

Table 1. Summary of Dataset

| Dataset | # Attributes | # instances | #defected | Language |
|---|---|---|---|---|
| JM1 | 22 | 10855 | 80.65% | C |
| CM1 | 22 | 498 | 9.83% | C |
| KC1 | 22 | 522 | 20.5% | C++ |
| PC1 | 22 | 1109 | 93.05% | C |
| Class-level KC1 ver1 | 95 | 145 | - | C and C++ |
| Class-level KC1 ver2 | 95 | 145 | - | C and C++ |

Table 2. The summary of code metrics

| Number | Quality metrics | Description |
|---|---|---|
| 1 | loc (v) | line count of code |
| 2 | v (g) | Cyclomatic complexity |
| 3 | ev (g) | Essential complexity |
| 4 | iv (g) | Design complexity |
| 5 | loCode | line count |
| 6 | loComment | Count of lines of comments |
| 7 | loBlank | Count of blank lines |
| 8 | loCodeAndComment | Count of code and comment lines |
| 9 | uniq_Op | Unique operators |
| 10 | uniq_Opnd | Unique operands |
| 11 | total_Op | Total operators |
| 12 | total_Opnd | Total operands |
| 13 | branchCount | Branch count of the flow graphs |
| 14 | n | total operators + operands |
| 15 | v | Volume |
| 16 | l | Program length |
| 17 | d | Difficulty |
| 18 | i | Intelligence |

| 19 | e | Effort |
|---|---|---|
| 20 | b | Estimate of the effort |
| 21 | t | Time estimator |
| 22 | Defect | True/False |

## 4. Research Methodology

In this paper, we will be exploring a solution to predict the defect in software using Support Vector Machine (SVM) with different kernel functions. The datasets that will be used are taken from NASA metrics Data Program, the number of features is 22 (4 McCabe metrics, 9 base Halstead measures, 8 derived Halstead measures and defect variable as output) as discussed before. Before using the Dataset, the Dataset will be pre-processed and cleaned by handling missing values and outliers. The datasets are divided to training and testing data. In Software Defect Predication (SDP) the selection of training data and testing data will be done in two different ways; the first one, in the same Dataset will be choosing the training and testing data randomly (or may be sequential). In second one, the training will be taking from Dataset as previous version and the testing data will be taking from another dataset as next version. We will use the first approach. The data will be handled and cleaned before running experiments. The proposed models will be validated using 10-cross validation. After that, the SVM with different kernel functions will be examined. The last step, the results will be compared and evaluated using classification accuracy measures such as: Recall, Precision, Classification Accuracy, and Balance. The tools that will be used are Rapid Miner for the implementation of our proposed solution. The accuracy of each model will be measured by the common accuracy measures: Recall, Precision, accuracy, Specificity, F-measure and Balance. Software Defect Prediction (SDP) detectors can be assessed according to confusion matrix or Error matrix: is a table used to describe the performance of classification model on a set of test data for which the true values are known. It is showed the number of correct and incorrect prediction, where is summarized with count values and broken down by each class. This is the key to the confusion matrix as shown in Table 3 [12].

Table 3. Confusion Matrix

|  | Predicted as defective | Predicted as non-defective |
|---|---|---|
| defective | TP | FN |
| Non defective | FP | TN |

Where TP is True positive which means correctly classified as defective module. TN is True negative which means correctly classified as non-defective module. FP is False positive which means classifies non defective module as defective module, and FN is False negative which means classifies defective module as non-defective module.

To correctly identify a defective prediction, the "Precision" is used to determine the defective prediction rate, or the extent of the prediction is originally defective, or not. Recall is also called sensitivity, probability of detection (pd), or true positive rate (TPR). There are also many measures called probability of false alarm (pf) or false positive rate (FPR) which suggests the percentage of false defective predictions. Based on what has already, an optimal predictor should achieve TPR (pd) is 1, FPR (pf) is 0 and the Precision is 1. When the TPR and FPR are plotted, the result in Receiver Operating Characteristics (ROC) curve and from ROC the area under the curve (AUC) is to be measured. AUC is measured between 0 and 1, with 1 being the optimal solution point. Table 4 presents performance measures [12]. The, the data must be cleaned from missing value and outliers. The existing of missing values and outliers hinder the success of building accurate learning models therefore researchers suggested using some statistical tool to ignore these outliers such as boxplot. The missing values can be handled by either replacing them with the feature average of ignoring them. In this paper we ignore missing values because they are a few. The proposed algorithm must be validated using robust validation procedure such as cross validation and bootstrapping. During validation procedure the data is divided into training and testing subsets and training data is entered to learning the model while the testing data is used to evaluate accuracy of the model.

## 5. Experimental Results

This section presents the results of the experiment study, which has been conducted to validate our module. The evaluation has been performed on Support Vector Machine (SVM) with different Kernel functions, using public datasets obtained from PROMISE data repository as described in Dataset section. To evaluate the performance of each proposed model, used 10-Folds cross-validation approach. This procedure divides the datasets randomly into 10-fold equal size subsets, where in each fold 9 subsets are used for training and one subset is used for testing. This process is repeated 10 times until all subsets act as testing data as described in section 3. In each experiment SVM model with different kernel function is constructed under two perspectives: using all features and using feature subset selected by PCA technique. Furthermore, six kernel functions were used: Linear, Quadratic, Cubic, Gaussian, RBF, Sigmoid.

Table 4. Performance measures

| metric | Definition of the measure |
|---|---|
| Sensitivity | $\frac{TP}{TP + FN}$ |
| Precision | $\frac{TP}{TP + FP}$ |
| False positive rate | $\frac{FP}{FP + TN}$ |
| Specificity | $\frac{TN}{TN + FP}$ |
| Accuracy | $\frac{TN + TP}{TN + FN + TP + FP}$ |
| Balanced Accuracy | $1 - \frac{\sqrt{(0 - pf)^2 + (1 - pd)^2}}{\sqrt{2}}$ |

## 5.1 CM1 Dataset Result

It can be noted from Table 5 that the Recall and Precision values are unacceptable for all kernel functions because their values are close to zero. Specificity values are very good for all kernel functions, with relatively similar values. Balance values are not very bad with a range between (0.29 - 0.4). Accuracy values are very good, as almost 90% of all kernel functions are good. TRP and FPR values are unacceptable for all kernel functions because they are nearly zero. "Area Under Curve" is acceptable for all kernel functions ranging from (0.50 - 0.64). With respect to all performance results, better solutions are observed for the Quadratic kernel function than the other five kernel functions with all features in the CM1 Dataset.

It can be noted from Table 6 that the Recall and Precision values are unacceptable for all kernel functions; specificity values are very good for all kernel functions, with similar values. Balance values are not very bad with a range between (0.29 - 0.4). For all kernel functions with a range between (0.86 - 0.95) the accuracy values are so good. TRP and FPR values are unacceptable for all kernel functions, because they're almost zero. With all kernel functions with a range between (0.50 - 0.71) the values "Area Under Curve" are acceptable. With respect to all performance results, better solutions are observed for the Quadratic kernel function than the other five kernel functions considered with selected features in the CM1 Dataset. It was little improvement in all performance results when using selected features in CM1 Dataset. This is because of the features selected that were used.

## 5.2 KC1 Dataset Result

From Table 7 we can note that for all kernel functions the Recall values and Precision values are acceptable. Specificity values are generally good, as they are almost 96% for all kernel functions, with the exception for Sigmoid kernel that obtained of the 86%. Balance values with a range of (0.32 -0.55) are fairly good. The accuracy values for all kernel functions are relatively good with range between (0.78 - 0.84); TRP values are acceptable for all kernel functions except for the Cubic kernel function. FPR values for all kernel functions are unacceptable, as they are almost zero. "Area Under Curve" values for all kernel functions with a range between (0.66 - 0.81) are acceptable. With respect to all performance results, better solutions are observed for the RBF kernel function than the other five kernel functions considered with all features in the KC1 Dataset.

Table 5. Performance results of the SVM kernel functions on CM1 Datasets, using all Features.

| Kernel | Recall | Precision | Specificity | Balance | Accuracy | TPR | FPR | AUC |
|---|---|---|---|---|---|---|---|---|
| Linear | 0.00 | 0.00 | 1.00 | 0.29 | 0.90 | 0.00 | 0.00 | 0.62 |
| Quadratic | 0.16 | 0.42 | 0.98 | 0.41 | 0.90 | 0.16 | 0.02 | 0.64 |
| Cubic | 0.16 | 0.22 | 0.94 | 0.41 | 0.86 | 0.16 | 0.07 | 0.61 |

| Kernel | | | | | | | |
|---|---|---|---|---|---|---|---|
| Gaussian | 0.00 | 0.02 | 1.00 | 0.29 | 0.90 | 0.00 | 0.00 | 0.57 |
| RBF | 0.00 | 0.00 | 1.00 | 0.29 | 0.90 | 0.00 | 0.00 | 0.50 |
| Sigmoid | 0.08 | 0.29 | 0.98 | 0.35 | 0.89 | 0.08 | 0.00 | 0.53 |

From Table 8 we can note that the values Recall and Precision are acceptable for all functions of the kernel. All kernel functions have very good specificity values, with a range between (0.85 - 0.97). Balance values with a range between (0.31 - 0.49) are acceptable. Accuracy values are so good for all kernel functions; as they are close to 84% except for RBF is 77%. For all kernel functions, TRP values are acceptable; FPR values are unacceptable, as for all kernel functions they are almost at zero.

Table 6. Performance results of SVM kernel functions on CM1 Datasets, using PCA

| Kernel | Recall | Precision | Specificity | Balance | Accuracy | TPR | FPR | AUC |
|---|---|---|---|---|---|---|---|---|
| Linear | 0.00 | 0.00 | 1.00 | 0.29 | 0.90 | 0.00 | 0.00 | 0.44 |
| Quadratic | 0.18 | 0.41 | 0.97 | 0.42 | 0.89 | 0.18 | 0.03 | 0.71 |
| Cubic | 0.20 | 0.28 | 0.94 | 0.44 | 0.90 | 0.00 | 0.00 | 0.66 |
| Gaussian | 0.00 | 0.03 | 1.00 | 0.29 | 0.90 | 0.00 | 0.00 | 0.63 |
| RBF | 0.00 | 0.00 | 1.00 | 0.29 | 0.95 | 0.00 | 0.00 | 0.50 |
| Sigmoid | 0.10 | 0.19 | 0.95 | 0.36 | 0.87 | 0.10 | 0.05 | 0.53 |

For all kernel functions with a range between (0.65 - 0.83) the values "Area Under Curve" are acceptable. With respect to all performance results, better solutions are observed for the Quadratic kernel function than the other five kernel functions considered with selected features in the KC1 Dataset.

When the selected features used in KC1 Dataset, all performance results were not improved except for Area Under Curve. This is because of the selected features that have been used, and we do not know how the mechanism of selection entities in cross-validation.

### 5.3 PC1 Dataset Result

From Table 9 it can be noted that the Recall values for all kernel functions are Totally unacceptable. Precision values for all kernel functions are acceptable, except the value for the Sigmoid kernel function. Specificity values for all kernel functions are very good as they are close to 96 %. Balance values with a range of (0.31- 0.49) are acceptable. Accuracy values are good for all functions of the kernel; since they are close to 91%. TRP values are unacceptable, as they are almost zero for all kernel functions with the exception of the Cubic kernel. FPR values are insufficient for all kernel functions, because they are almost zero. "Area Under Curve" values for all kernel functions with a range between (0.53 - 0.73) are acceptable. With respect to all performance results, better solutions are observed for the Cubic kernel function than the other five kernel functions considered with all features in the PC1 Dataset.

From Table 10 it can be noted that the Recall values for all kernel functions are totally unacceptable. Except for the Sigmoid and Cubic kernel functions, precise values are acceptable for all kernel functions. Specificity values are very good for all kernel functions; with the exception of Cubic kernel function, they are close to 97%. Balance values with a range of (0.31 - 0.49) acceptable.

Table 7. Performance results of the SVM kernel functions on KC1 Datasets, using all Features.

| Kernel | Recall | Precision | Specificity | Balance | Accuracy | TPR | FPR | AUC |
|---|---|---|---|---|---|---|---|---|
| Linear | 0.36 | 0.75 | 0.97 | 0.54 | 0.84 | 0.36 | 0.03 | 0.81 |
| Quadratic | 0.37 | 0.69 | 0.96 | 0.56 | 0.84 | 0.37 | 0.04 | 0.73 |
| Cubic | 0.37 | 0.54 | 0.97 | 0.55 | 0.81 | 0.37 | 0.08 | 0.67 |
| Gaussian | 0.37 | 0.74 | 0.97 | 0.56 | 0.85 | 0.37 | 0.03 | 0.77 |
| RBF | 0.36 | 0.70 | 0.96 | 0.54 | 0.84 | 0.36 | 0.04 | 0.66 |

| | | | | | | | |
|---|---|---|---|---|---|---|---|
| Sigmoid | 0.46 | 0.46 | 0.86 | 0.33 | 0.78 | 0.46 | 0.14 | 0.66 |

Table 8. Performance results of the SVM kernel functions on KC1 Datasets using PCA

| Kernel | Recall | Precision | Specificity | Balance | Accuracy | TPR | FPR | AUC |
|---|---|---|---|---|---|---|---|---|
| Linear | 0.36 | 0.79 | 0.98 | 0.54 | 0.85 | 0.36 | 0.02 | 0.83 |
| Quadratic | 0.40 | 0.72 | 0.96 | 0.58 | 0.85 | 0.40 | 0.04 | 0.73 |
| Cubic | 0.43 | 0.62 | 0.93 | 0.59 | 0.83 | 0.43 | 0.07 | 0.65 |
| Gaussian | 0.36 | 0.68 | 0.96 | 0.55 | 0.84 | 0.36 | 0.04 | 0.69 |
| RBF | 0.35 | 0.66 | 0.95 | 0.54 | 0.83 | 0.35 | 0.05 | 0.65 |
| Sigmoid | 0.44 | 0.44 | 0.86 | 0.33 | 0.77 | 0.44 | 0.14 | 0.65 |

For all kernel functions, accuracy values are so good; with a range between (0.90 - 0.93) except for the Cubic kernel function, it is 77%. TRP and FPR values are unacceptable, because they are almost zero for all functions of the kernel except for the Cubic kernel. "Area Under Curve" values for all kernel functions with a range between (0.51 - 0.75) are acceptable. With respect to all performance results, better solutions are observed for the Gaussian kernel function than the other five kernel functions considered with selected features in the PC1 Dataset. There was no improvement in all performance results except in accuracy when we used selected features in the PC1 Dataset. This is due to a function of the kernel, the selected features that were used and we don't know how the process of selection entities in cross-validation.

Table 9. Performance results of the SVM kernel functions on PC1 Datasets using all Features.

| Kernel | Recall | Precision | Specificity | Balance | Accuracy | TPR | FPR | AUC |
|---|---|---|---|---|---|---|---|---|
| Linear | 0.03 | 0.67 | 1.00 | 0.31 | 0.93 | 0.03 | 0.00 | 0.70 |
| Quadratic | 0.13 | 0.32 | 0.98 | 0.39 | 0.92 | 0.13 | 0.02 | 0.68 |
| Cubic | 0.29 | 0.39 | 0.97 | 0.49 | 0.92 | 0.29 | 0.03 | 0.67 |
| Gaussian | 0.08 | 0.55 | 1.00 | 0.35 | 0.93 | 0.08 | 0.01 | 0.73 |
| RBF | 0.08 | 0.67 | 1.00 | 0.35 | 0.93 | 0.08 | 0.00 | 0.54 |
| Sigmoid | 0.05 | 0.13 | 0.98 | 0.33 | 0.91 | 0.05 | 0.03 | 0.53 |

Table 10. Performance results of the SVM kernel functions on PC1 Datasets using PCA

| Kernel | Recall | Precision | Specificity | Balance | Accuracy | TPR | FPR | AUC |
|---|---|---|---|---|---|---|---|---|
| Linear | 0.20 | 0.71 | 0.98 | 0.42 | 0.95 | 0.20 | 0.00 | 0.75 |
| Quadratic | 0.12 | 0.43 | 0.99 | 0.38 | 0.93 | 0.12 | 0.01 | 0.63 |
| Cubic | 0.40 | 0.10 | 0.83 | 0.50 | 0.75 | 0.48 | 0.22 | 0.55 |
| Gaussian | 0.10 | 0.67 | 1.00 | 0.35 | 0.93 | 0.13 | 0.00 | 0.69 |
| RBF | 0.08 | 0.72 | 1.00 | 0.37 | 0.93 | 0.11 | 0.00 | 0.53 |
| Sigmoid | 0.07 | 0.16 | 0.97 | 0.35 | 0.91 | 0.07 | 0.03 | 0.51 |

### 5.4 JM1 Dataset Result

It can be noted from Table 11 that the Recall values are unacceptable for all kernel functions except for the Cubic kernel function; it is 77%. Precision values are acceptable for all functions of the kernel except the functions Sigmoid and RBF kernel. Specificity values are very good, as they are similar to one% of Linear, Quadratic, and Gaussian kernel functions, becoming nearly 90 % of RBF kernel functions and unacceptable for Cubic and Sigmoid kernel functions. Balance values with a range of (0.21 - 0.36) are acceptable. For linear, quadratic, and Gaussian kernel functions, accuracy values are so good; as they are close to 81 percent, but unacceptable in the functions of Cubic, RBF, and Sigmoid. For all kernel functions except the Cubic kernel function, TRP and FPR values are unacceptable because they are nearly zero. "Area Under Curve" values are acceptable for all kernel functions with a range between (0.51 - 0.75). expect for cubic kernel function. With respect to all performance results, better solutions are observed for the Gaussian kernel function than the other five kernel functions with all features considered in the JM1 Dataset.

It can be noted from Table 12 that the Recall values are unacceptable for all kernel functions except for the Cubic and Quadratic kernel functions, 40% and 61% are in order. Precision values are acceptable for all functions of the kernel, with the exception of Cubic and Quadratic functions. Specificity value is very good, as it is close to one for Linear, RBF, and

Gaussian kernel functions, as it is close to 82% for Sigmoid kernel function and unacceptable for Cubic and Quadratic kernel functions. Balance values with a range of (0.29 - 0.49) are not that bad. Accuracy values are so good for linear, RBF kernel functions as they are nearly 81%. In Sigmoid kernel function is 77% unacceptable in Cubic, Quadratic, and Gaussian kernel function. TRP and FPR values are unacceptable for all kernel functions except for the Cubic and Quadratic kernel functions, as they are almost zero. "Area Under Curve" values are acceptable for all kernel functions with a range between (0.50 - 0.63) expect for Quadratic kernel function. With respect to all performance results, better solutions are observed for the RBF kernel function than the other five kernel functions considered with selected features in the JM1 Dataset. It was improvement when selected features used in JM1 Dataset, and no improvement in all performance. This is due to a function of the kernel, the selected features that were used and we don't know how the process of selection entities in cross-validation.

Table 11. Performance results of the SVM kernel functions on JM1 Datasets using all Features.

| Kernel | Recall | Precision | Specificity | Balance | Accuracy | TPR | FPR | AUC |
|---|---|---|---|---|---|---|---|---|
| Linear | 0.02 | 0.71 | 1.00 | 0.31 | 0.81 | 0.02 | 0.00 | 0.65 |
| Quadratic | 0.09 | 0.54 | 0.98 | 0.36 | 0.81 | 0.09 | 0.02 | 0.64 |
| Cubic | 0.76 | 0.19 | 0.21 | 0.21 | 0.32 | 0.76 | 0.79 | 0.48 |
| Gaussian | 0.10 | 0.61 | 0.99 | 0.37 | 0.81 | 0.10 | 0.02 | 0.62 |
| RBF | 0.01 | 0.34 | 0.90 | 0.30 | 0.18 | 0.01 | 0.10 | 0.55 |
| Sigmoid | 0.06 | 0.28 | 0.59 | 0.27 | 0.19 | 0.06 | 0.41 | 0.54 |

## 5.5 Class-Level Data for KC1 version 1 Dataset Result

From Table 13, it can be noted that the Recall values are acceptable for all kernel functions except the RBF and sigmoid kernel functions as they are nearly zero. Precision values are acceptable for all kernel functions except the RBF and sigmoid kernel functions, as they are nearly to zero. Specificity values are very good, as they are near to one for RBF and Sigmoid kernel functions, other Kernels with rang between (0.66 -0 .83).

Table 12. Performance results of the SVM kernel functions on JM1 Datasets using PCA

| Kernel | Recall | Precision | Specificity | Balance | Accuracy | TPR | FPR | AUC |
|---|---|---|---|---|---|---|---|---|
| Linear | 0.01 | 0.63 | 1.00 | 0.30 | 0.81 | 0.01 | 0.00 | 0.63 |
| Quadratic | 0.62 | 0.18 | 0.32 | 0.45 | 0.38 | 0.62 | 0.68 | 0.45 |
| Cubic | 0.41 | 0.20 | 0.60 | 0.49 | 0.56 | 0.41 | 0.40 | 0.50 |
| Gaussian | 0.09 | 0.60 | 0.96 | 0.35 | 0.58 | 0.09 | 0.04 | 0.59 |
| RBF | 0.08 | 0.58 | 0.99 | 0.35 | 0.81 | 0.08 | 0.01 | 0.53 |
| Sigmoid | 0.34 | 0.32 | 0.83 | 0.52 | 0.73 | 0.34 | 0.17 | 0.58 |

Balance values are very good, as they are nearly to 77% except in RBF and Sigmoid kernel functions. Accuracy values are good as they are nearly to 77% in all kernel functions expect unacceptable in RBF, Sigmoid kernel functions. TRP values are good as they are nearly to 80% for all kernel functions except for the Sigmoid and RBF kernel functions. FBR values are unacceptable as they are nearly to zero for all kernel functions except for the Cubic, Linear and Quadratic kernel functions. "Area Under Curve" values are acceptable within the range between (0.50 - 0.84) for all kernel functions. For all performance results, better solutions are observed for the Gaussian kernel function than the other five considered kernel functions in the KC1version 1 dataset class-level data with all features.

Table 13. Performance results of the SVM kernel functions on Class-level data for KC1version 1 Datasets, Using all Features.

| Kernel | Recall | Precision | Specificity | Balance | Accuracy | TPR | FPR | AUC |
|---|---|---|---|---|---|---|---|---|
| Linear | 0.82 | 0.81 | 0.73 | 0.77 | 0.79 | 0.82 | 0.27 | 0.84 |
| Quadratic | 0.85 | 0.78 | 0.67 | 0.74 | 0.77 | 0.85 | 0.33 | 0.81 |
| Cubic | 0.79 | 0.77 | 0.67 | 0.72 | 0.74 | 0.79 | 0.33 | 0.79 |
| Gaussian | 0.75 | 0.87 | 0.83 | 0.79 | 0.79 | 0.75 | 0.02 | 0.83 |
| RBF | 0.00 | 0.00 | 1.00 | 0.00 | 0.43 | 0.00 | 0.00 | 0.50 |

|  | Sigmoid | 0.00 | 0.00 | 0.97 | 0.29 | 0.00 | 0.03 | 0.41 | 0.51 |

Table 14. Performance results of the SVM kernel functions on Class-level data for KC1version 1 Datasets using PCA

| Kernel | Recall | Precision | Specificity | Balance | Accuracy | TPR | FPR | AUC |
|---|---|---|---|---|---|---|---|---|
| Linear | 0.84 | 0.70 | 0.48 | 0.62 | 0.39 | 0.84 | 0.52 | 0.77 |
| Quadratic | 0.78 | 0.70 | 0.52 | 0.62 | 0.67 | 0.78 | 0.48 | 0.67 |
| Cubic | 0.71 | 0.71 | 0.58 | 0.64 | 0.66 | 0.71 | 0.42 | 0.65 |
| Gaussian | 0.90 | 0.69 | 0.40 | 0.57 | 0.70 | 0.90 | 0.60 | 0.08 |
| RBF | 0.70 | 0.65 | 0.73 | 0.71 | 0.72 | 0.70 | 0.27 | 0.71 |
| Sigmoid | 0.43 | 0.68 | 0.86 | 0.59 | 0.68 | 0.43 | 0.14 | 0.65 |

From Table 14, it can be noted that for all kernel functions the Recall values and Precision values are acceptable. For all kernel functions, specificity values are unacceptable, except for functions in the Sigmoid and RBF kernels. Balance values with a range of (0.57 - 0.71) are very good. Accuracy values are good for all kernel functions because they are nearly 66%. All kernel functions are good at the TRP and FBR values. "Area Under Curve" values are acceptable with a range of (0.65 - 0.80) for all kernel functions. For all performance results, better solutions are observed for the Gaussian kernel function than the other five considered kernel functions in the KC1version 1 dataset class-level data with selected features. Dataset used in class-level data were improved in Recall, Precision, Balance, TPR and FPR, there was no improvement in Area Under Curve and another performance was getting bad including Specificity and Accuracy. It is due to a function of the kernel, the selected features that were used and we don't know how the mechanism of selection entities in cross-validation.

## 5.6 Class-Level Data for KC1 version 2 Dataset Result

From Table 15, it can be noted that for all kernel functions, the Recall values are nearly 100%, except for the sigmoid and RBF kernel functions, as they are close to zero. Precision values are so good because the majority values are close to 95% in order for all kernel functions except RBF and sigmoid kernel functions. Specificity values are unacceptable because for all kernel functions, they are close to zero except for the Sigmoid and RBF kernels; they are 100% and 87% in order. Balance values with a range between (0.29-0.38) are not that bad, except for functions in the RBF and Sigmoid kernel, because they are nearly zero. The accuracy values are so good that they are near to 94 percent for all kernel functions except RBF, that it's nearly zero. TRP and FPR values are as good as similar to one for all kernel functions with the exception of the RBF and Sigmoid kernel functions, as they are nearly zero. "Area Under Curve" values are acceptable with a range between (0.50 - 0.74) Cubic, Sigmoid, and RBF kernel function, and other kernels are near to one. With respect to all performance results, better solutions are observed for the Gaussian kernel function than the other five kernel functions considered in the KC1version 2 dataset class-level data with all features.

From Table 16, it can be noted that the Recall values are nearly to one for all kernel functions except the sigmoid and RBF kernel functions, as they are near to zero. Precision values are so good that for all kernel functions, except for RBF and sigmoid kernel functions, they are close to 97%,
because they are nearly zero. Specificity values are unacceptable because they are nearly zero for all kernel functions, except for the Sigmoid and RBF kernels, as they are near to one, and in the Cubic kernel function, they are 50%. With a range of (0.29-0.47), balance values are acceptable except for Cubic and Quadratic kernel functions, as they are close to 64 %. TRP values are as good as near to one for all kernel functions except for the RBF and Sigmoid kernel functions as near to zero%. With the exception of the RBF and Sigmoid kernel functions, FPR values are so good with a range of (0.50 - 1.0) because they are close to zero %. "Area Under Curve" values are acceptable for the feature of the Sigmoid and RBF kernels, as they are nearly 50% other kernels, they are almost 85%. For all performance results, better solutions are observed for the Gaussian kernel function than the other five kernel functions considered in the KC1version 2 dataset class-level data with all features.

The comparison of the proposed classifier with the Support Vector Machine (SVM) with different kernel functions applied for the same NASA datasets in terms of the performance metrics: sensitivity, specificity, probability of false alarm, balance, accuracy, and area under the curve. The results were different within the Dataset because each Dataset has a different number of entities, some data have 125 such as KC1version 2 class-level data, other datasets have 10,000 entities and this affects the cross-validation selection process. For some datasets the selected features perform better in the same

Dataset and in other datasets there is no improvement at all.

Table 15. Performance results of the SVM kernel functions on Class-level data for KC1version 2 Datasets using all Features.

| Kernel | Recall | Precision | Specificity | Balance | Accuracy | TPR | FPR | AUC |
|---|---|---|---|---|---|---|---|---|
| Linear | 0.99 | 0.95 | 0.13 | 0.38 | 0.95 | 0.99 | 0.88 | 0.93 |
| Quadratic | 0.99 | 0.95 | 0.13 | 0.38 | 0.95 | 0.99 | 0.88 | 0.92 |
| Cubic | 0.97 | 0.95 | 0.13 | 0.38 | 0.92 | 0.97 | 0.88 | 0.74 |
| Gaussian | 1.00 | 0.95 | 0.00 | 0.00 | 0.95 | 1.00 | 1.00 | 0.89 |
| RBF | 0.00 | 0.00 | 1.00 | 0.29 | 0.95 | 0.00 | 0.00 | 0.50 |
| Sigmoid | 0.01 | 0.50 | 0.88 | 0.08 | 0.06 | 0.01 | 0.13 | 0.51 |

Table 16. Performance results of the proposed predictor on Class-level data for KC1version 2 Datasets using PCA

| Kernel | Recall | Precision | Specificity | Balance | Accuracy | TPR | FPR | AUC |
|---|---|---|---|---|---|---|---|---|
| Linear | 0.99 | 0.96 | 0.25 | 0.47 | 0.95 | 0.99 | 0.75 | 0.81 |
| Quadratic | 0.99 | 0.97 | 0.50 | 0.65 | 0.96 | 0.99 | 0.50 | 0.85 |
| Cubic | 0.99 | 0.97 | 0.38 | 0.64 | 0.96 | 0.99 | 0.63 | 0.83 |
| Gaussian | 1.00 | 0.95 | 0.00 | 0.29 | 0.95 | 1.00 | 1.00 | 0.83 |
| RBF | 0.00 | 0.00 | 1.00 | 0.29 | 0.95 | 0.00 | 0.00 | 0.50 |
| Sigmoid | 0.00 | 0.00 | 0.99 | 0.29 | 0.94 | 0.00 | 0.01 | 0.50 |

# 6. CONCLUSION

Software Defect Prediction is a vital task during software development to help testing team to focus on defect proneness modules. To support that, various machine learning methods have been used to build models that can predict faulty modules based on datasets collected from software industries. Among them, Support vector machine has shown good performance for this problem, but there are no prior studies examined the performance of kernel functions for defect prediction problem. Thus, this research we will examine the performance of support vector machine with different kernel functions over different datasets collected from software data repositories. The results demonstrate that there is no kernel function that can give stable performance across different experimental settings. In addition, the use of feature subset selection using PCA did improve accuracy of kernel functions over some datasets. In CM1 Dataset, better solutions are observed for the Quadratic kernel function than the other five kernel functions with all and selection features. In KC1 Dataset, better solutions are observed for the RBF kernel function than the other five kernel functions with all and selection features. In PC1 Dataset, better solutions are observed for the Cubic kernel function than the other five kernel functions considered with all features, but when select some features, the better solutions are observed for the Gaussian kernel function than the other five kernel functions. In JM1 Dataset, better solutions are observed for the Gaussian kernel function than the other five kernel functions with all features, but when select some features, better solutions are observed for the RBF kernel function than the other five kernel functions. In Class-level data for KC1version 1 dataset, better solutions are observed for the Gaussian kernel function than the other five considered kernel functions in all and selected features. In in the KC1version 2 dataset class-level Dataset better solutions are observed for the Gaussian kernel function than the other five kernel functions considered with all and selected features. The results were different within the Dataset because each Dataset has a different number of entities, some data have 125 such as KC1version 2 class-level data, other datasets have 10,000 entities and this affects the cross-validation selection process. For some datasets the selected features perform better in the same Dataset and in other datasets there is no improvement at all.